\documentclass[twocolumn,aps,prl,showpacs,preprintnumbers,amsmath,amssymb,floatfix,reprint]{revtex4}
\usepackage{bm}% bold math
\usepackage{amsmath}
\usepackage{graphicx}% Include figure files
\usepackage{dcolumn}% Align table columns on decimal point
\usepackage[mathcal]{euscript}

\def\beq{\begin{equation}}
\def\eeq{\end{equation}}

\usepackage[amssymb, cdot, thinqspace]{SIunits}
\usepackage{nicefrac}

\begin{document}

\title{High-Precision Measurement of Rydberg State Hyperfine Splitting in a Room-Temperature Vapour Cell}

\author{Atreju Tauschinsky}
\email{Atreju.Tauschinsky@uva.nl}
\author{Richard Newell}
\author{H. B. van Linden van den Heuvell}
\author{R. J. C. Spreeuw}
\email{spreeuw@science.uva.nl}
\affiliation{Van der Waals-Zeeman Institute, Institute of Physics, University of Amsterdam, \\ PO Box 94485, 1090 GL Amsterdam, The Netherlands}

\date{\today}

\begin{abstract}
We present direct measurements of the hyperfine splitting of Rydberg states in rubidium 87 using Electromagnetically Induced Transparency (EIT) spectroscopy in a room-temperature vapour cell. With this method, and in spite of Doppler-broadening, line-widths of \unit{3.7}{\mega\hertz} FWHM, i.e. significantly below the intermediate state natural linewidth are reached. This allows resolving hyperfine splittings for Rydberg s-states with $n=20\ldots24$. With this method we are able to determine Rydberg state hyperfine splittings with an accuracy of approximately \unit{100}{\kilo\hertz}. Ultimately our method allows accuracies of order \unit{5}{\kilo\hertz} to be reached. Furthermore we present a direct measurement of hyperfine-resolved Rydberg state Stark-shifts. These results will be of great value for future experiments relying on excellent knowledge of Rydberg-state energies and polarizabilities.
\end{abstract}

% pacs:
% 32.10.Fn	Fine and hyperfine structure
% 32.60.+i	Zeeman and Stark effects
% 32.80.Ee	Rydberg states
% 42.50.Gy	EIT

\maketitle

%\section{Introduction}
Rydberg atoms have recently received a great amount of attention, motivated by their large polarizabilities and strong dipole-dipole coupling. This interest is often stimulated by the suitability of Rydberg atoms to engineer long-range interactions for quantum information processing \cite{Jaksch2000, Lukin2001, Muller2009a} or the investigation of strongly correlated systems \cite{Henkel2010, Pohl2010, Schauß2012}. The research in ultra-cold Rydberg atoms has resulted in two landmark experiments demonstrating dipole-blockade for two individual atoms \cite{Urban2009, Gaetan2009}, but also further experiments on mesoscopic ensembles in the blockade regime \cite{Dudin2012}. Cold ensembles of Rydberg atoms have been used for electrometry \cite{Tauschinsky2010, Abel2011, Carter2011}. Electromagnetically Induced Transparency (EIT) has also been used to observe Rydberg dipole blockade in cold ensembles of atoms \cite{Pritchard2010a, Hofmann2012, Peyronel2012} and it has been proposed to directly observe dipole blockade using EIT \cite{Gunter2012a, Olmos2011}.

In addition, great progress has also been made exciting Rydberg atoms in room-temperature vapour cells. Indeed, coherent effects have been observed here as well \cite{Huber2011, Kuebler2010} and sensitive methods for electric field measurements in vapour cells \cite{Bason2010a}, as well as an alternative to EIT measurements \cite{Barredo2012} have been developed.

Excellent knowledge of the spectroscopy of Rydberg states both in the presence and absence of electric fields is crucial for all of these experiments. In particular, Rydberg hyperfine structure may limit the fidelity of quantum gates \cite{Walker2012} and undermine coherent evolution.
Here we show that high-precision hyperfine spectroscopy of rubidium Rydberg states is possible in a room-temperature vapour cell and investigate the hyperfine splitting for various Rydberg states. We also present hyperfine-resolved measurements of the Rydberg state polarizability. Previous measurements of the zero-field Rydberg state hyperfine splitting rely on mm-wave transitions in a magneto-optical trap, but the results are less precise than those presented here. No prior measurements of hyperfine-resolved Stark-shifts are known to us.

%\section{Experimental Setup}

\begin{figure}[ht]
	\centering
	\includegraphics[width=1.00\columnwidth]{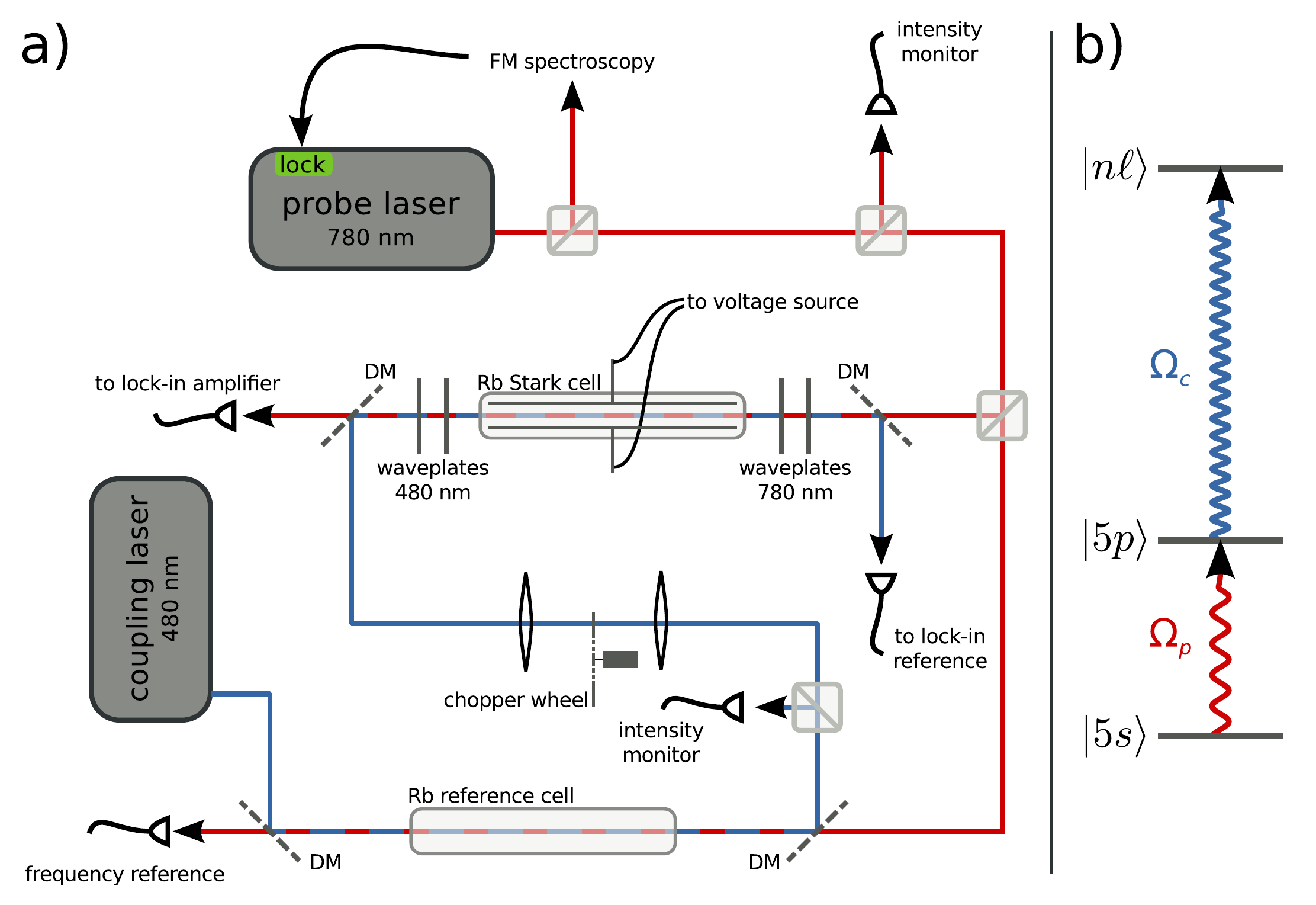}
	\caption{(Color online) (a) Schematic drawing of the setup used in the experiments. The probe laser is independently locked to a saturated absorption frequency-modulation (FM spectroscopy setup not shown here. The reference cell used to compensate long-term frequency drifts was only used for the Stark-map measurements shown in Fig. \ref{fig:starkshift}, but not for the hyperfine data presented in Figs. \ref{fig:trace} and \ref{fig:scaling}. DM: Dichroic Mirror. (b) Energy level diagram with the weak probe laser coupling the $5s$ ground to the $5p_{3/2}$ excited state with Rabi frequency $\Omega_p$ and the strong coupling laser connecting the excited state to a Rydberg state with Rabi frequency $\Omega_c$.}
	\label{fig:setup}
\end{figure}

A schematic drawing of the setup is shown in Fig. \ref{fig:setup}. At the heart of the experiment is a custom-made rubidium vapour cell. The cell is \unit{10}{\centi\meter} long and contains two internal stainless steel electrodes of \unit{95\times20}{\milli\meter\squared} size spaced \unit{5.35(3)}{\milli\meter} apart. The electrodes can be connected to a DC power supply and an Agilent 33250A function generator.

We perform EIT spectroscopy in this cell by counter-propagating a probe laser resonant with the $5s_{1/2}\rightarrow5p_{3/2}$ transition of $^{87}$Rb and a coupling laser coupling the $5p$ state to a Rydberg state through the cell. The probe laser is derived from a Toptica DL-100 external cavity diode laser at \unit{780.24}{\nano\meter} frequency-stabilized by saturated-absorption frequency-modulation (FM) spectroscopy in a separate vapour cell to the F=2 to F$'$=2 hyperfine transition. The coupling laser is derived from a frequency-doubled amplified diode laser system (Toptica TA-SHG Pro) at $\approx$\,\unit{480}{\nano\meter} and scanned across a Rydberg resonance. Both lasers propagate through the cell parallel to the long axis of the field plates and are overlapped over the entire length of the cell. The gaussian beam waists are approximately \unit{0.4}{\milli\meter} for the probe and \unit{1.0}{\milli\meter} for the coupling lasers, with peak intensities of \unit{0.4}{\milli\watt\per\centi\meter\squared} and \unit{4.3}{\milli\watt\per\centi\meter\squared} respectively.

The coupling laser is modulated by a chopper wheel at approximately \unit{4}{\kilo\hertz} for lock-in detection of the EIT signal. A  reference vapour cell without electric field plates is used to measure and compensate for drifts of the coupling laser frequency during scans.

\begin{figure}[htb]
	\centering
	\includegraphics[width=1.00\columnwidth]{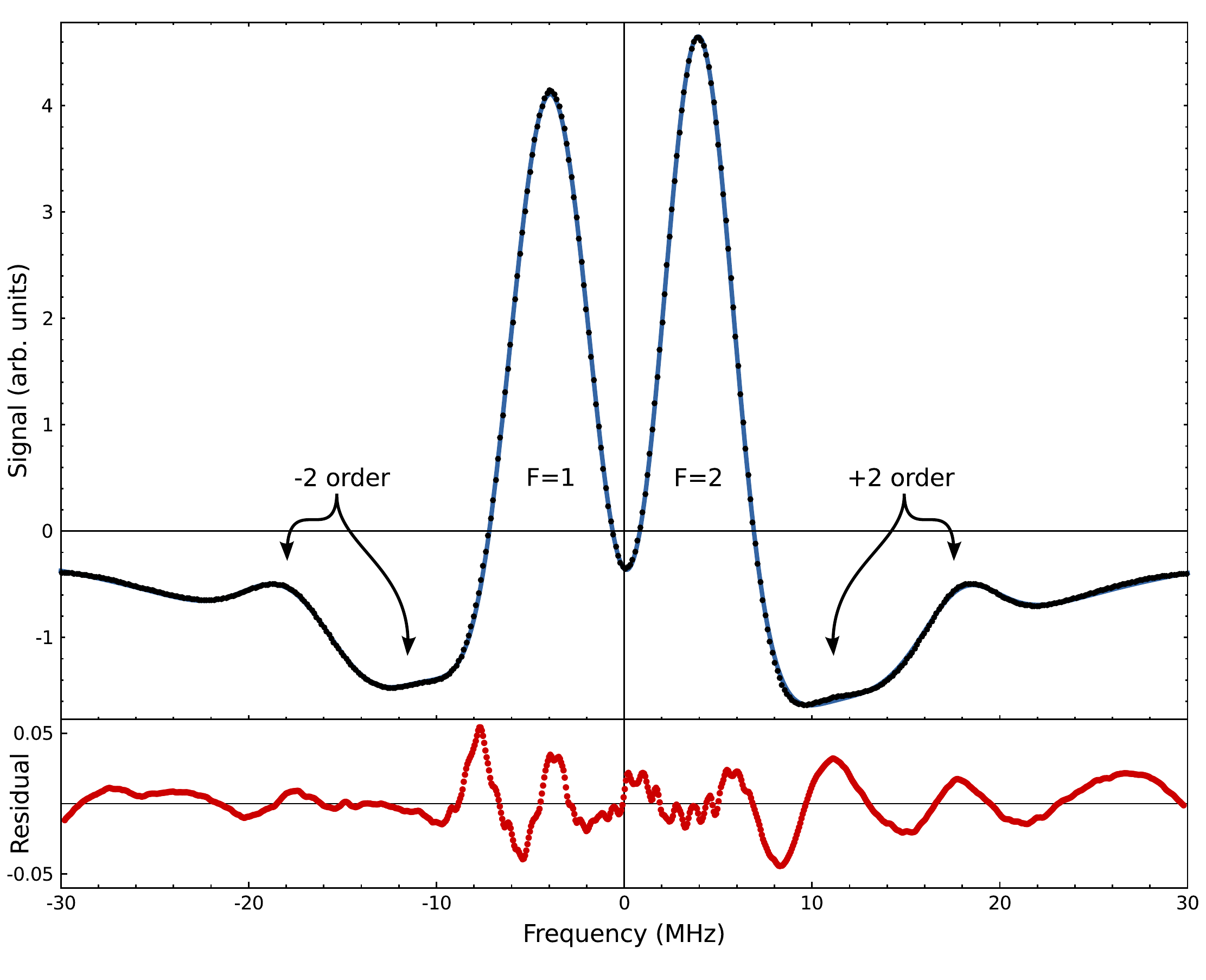}
	\caption{(Color online) Spectrum of 20s hyperfine structure including positive and negative 2nd order sidebands used for frequency calibration. Blue dots: average of 860 traces; Light-blue line: fit based on equation \eqref{eq:chi}. The lower part of the figure shows the residual of this fit. The field was modulated at \unit{7}{\mega\hertz}. First-order sidebands are not visible as their excitation is dipole-forbidden.}
	\label{fig:trace}
\end{figure}

%\section{Hyperfine States}
We measure Rydberg state hyperfine splittings by scanning the coupling laser across a Rydberg resonance by turning the grating of the ECDL head of the SHG system with the built-in piezo-element. The resulting spectrum is shown in Fig. \ref{fig:trace} for the $20s$-state. The frequency axis is calibrated by applying a \unit{7}{\mega\hertz} sinusoidally varying voltage to the field plates of the vapour cell, thereby creating sidebands of the state at a well-defined frequency spacing.

We assume the weak-probe limit, i.e. the probe Rabi frequency $\Omega_p \rightarrow 0$, where a model of the form
\begin{equation}
\chi \propto \int_{-\infty}^{\infty}\frac{i}{\gamma_p - i \Delta_p +\frac{\nicefrac{\Omega_c^2}{4}}{\gamma_c - i\left(\Delta_p+\Delta_c\right)}}N(v)dv
\label{eq:chi}
\end{equation}
for the susceptibility $\chi$ of the probe transition is valid \cite{Gea-Banacloche1995}. Here $\Omega_c$ is the coupling Rabi frequency, the probe and coupling detunings depend on the velocity of the atoms through Doppler shifts:
\begin{align*}
\Delta_p &= \Delta_p^0 - k_p v\\
\Delta_c &= \Delta_c^0 + k_c v
\end{align*}
and $\gamma_p$ and $\gamma_c$ are decay rates given by $\gamma_p = \nicefrac{1}{2}\Gamma_{5p}$ with $\Gamma_{5p}$ the natural decay rate of the excited state and $\gamma_c = \nicefrac{1}{2}\Gamma_{r}$ with $\Gamma_{r}$ the natural decay rate of the Rydberg state. Additional broadening effects can be included in $\gamma_c$ (see below).
$N(v)$ is a one-dimensional Maxwell-Boltzmann velocity distribution describing the velocity of the atoms in the vapour cell. The integral over $v$ is equivalent to the averaging over all velocity classes that occurs in a room-temperature vapour cell and can be solved analytically for a Maxwell-Boltzmann velocity distribution \cite{Gea-Banacloche1995}. 

The imaginary part of $\chi$ determines the absorption of the probe laser. We fit the data to an incoherent sum of six peaks of the form \eqref{eq:chi} after analytic integration as in \cite{Gea-Banacloche1995}. Each of the two hyperfine peaks is fitted with an independent coupling Rabi frequency, but sidebands share the coupling Rabi frequency of the main peaks. The intermediate state linewidth is fixed to the literature value, $\gamma_p = \unit{2\pi\times 3.03}{\mega\hertz}$ \cite{Steck2010}. The excited state linewidth $\gamma_c$ is fitted to a common value for all peaks. The fixed separation of the sidebands allows us to precisely calibrate the frequency axis and thus extract accurate values for the hyperfine splitting from our data. 

Fig. \ref{fig:trace} shows a spectrum for a Rydberg state $20s$ and the corresponding fit. The data points are an average of 860 individual traces, aligned by fitting two gaussians to the main hyperfine peaks and centering the midpoint between the peaks before averaging. Data and fit are virtually indistinguishable, confirming the quality of our measurements. The linewidth of our features is particularly remarkable: The fitted $\gamma_c$ is typically \unit{2\pi\times 2}{\mega\hertz}, significantly smaller than the intermediate state linewidth, even though all measurements are done in a vapour cell at room temperature, and with a free-running coupling laser. The observed width of a single hyperfine peak of about \unit{2\pi\times 3.7}{\mega\hertz} FWHM is however still somewhat larger than the limit of \unit{2\pi\times 1.7}{\mega\hertz} for vanishing $\gamma_c$ and $\Omega_c$ that can be observed in rubidium at room temperature. This is due to both the finite linewidth of the free-running coupling laser system as well as transit-time broadening due to atoms mving in and out of the beam radially \cite{Thomas1980}, which we estimate at approximately \unit{1}{\mega\hertz} for our beam width.

We perform similar measurements for Rydberg s states with principal quantum numbers $n=20\ldots24$. At $n>24$ the Doppler-broadened linewidth of the EIT resonance is too large to observe individual peaks. At $n<20$ the spectral tuning range of our laser system is limited. 
\begin{figure}[htb]
	\centering
	\includegraphics[width=1.00\columnwidth]{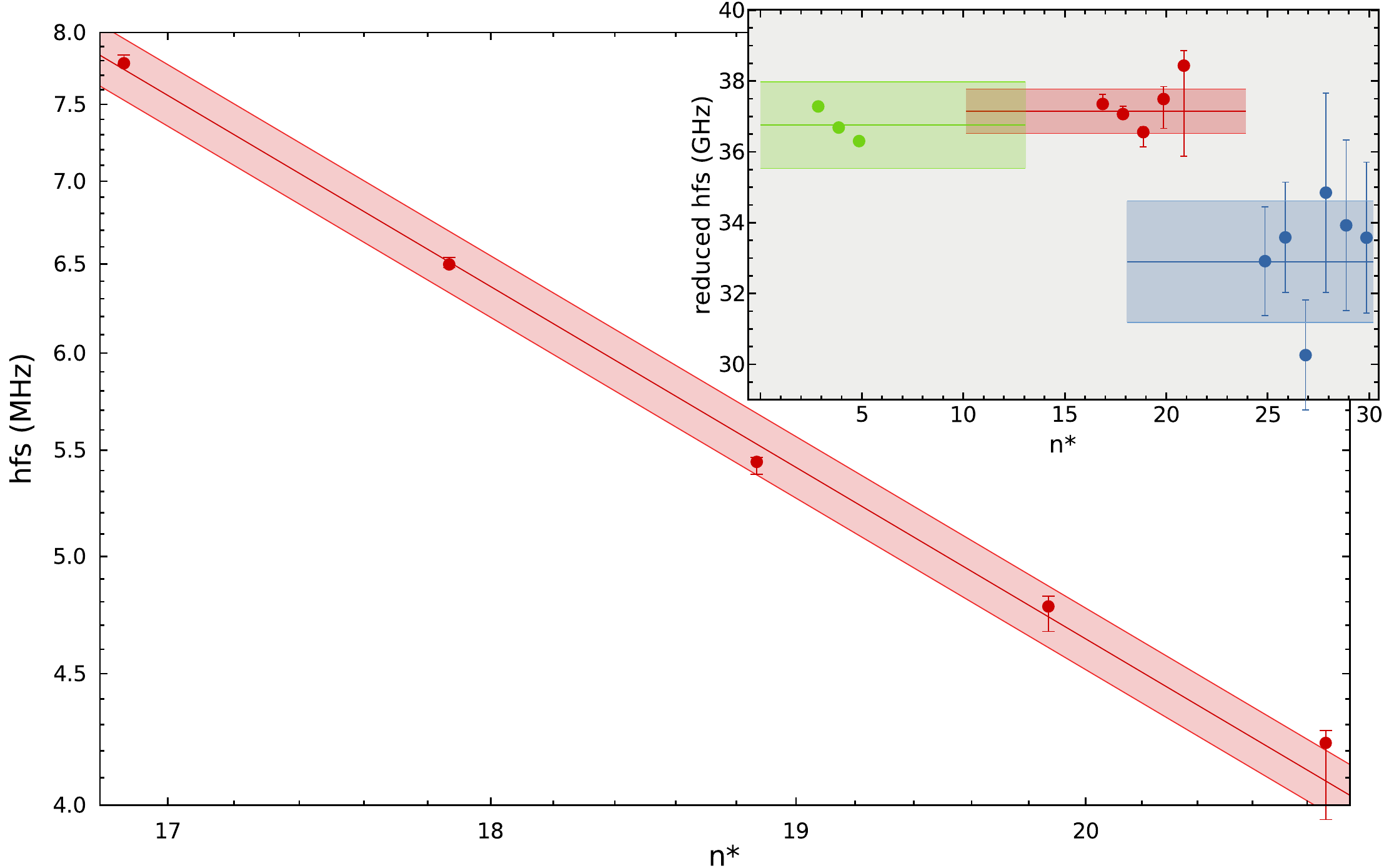}
	\caption{(Color online) Scaling of the hyperfine splitting with effective principal quantum number $n^*=(n-\delta)$, extracted from measurements such as presented in Fig. \ref{fig:trace}. The solid line is based on a $(n-\delta)^{-3}$-scaling with only the pre-factor as fit parameter. The shaded area signifies the  $95\%$ confidence region of this fit. The inset shows the same data after removing the $(n-\delta)^{-3}$ scaling in comparison to low-n data from reference \cite{Corney1977} and slightly higher-lying states from reference \cite{Li2003}. Error bars indicated are estimated on the basis of piezo scan nonlinearity, see text.}
	\label{fig:scaling}
\end{figure}

\begin{table}
\begin{ruledtabular}
\begin{tabular}{ccccc}
$n$ & $\nu_\mathrm{hfs}$ & $\sigma_\mathrm{fit}$ & $\sigma_\mathrm{piezo}$ & $\Delta_\mathrm{scaling}$ \\\hline
20 & 7 782 & 4 &  $\nicefrac{+57}{-17}$ & -43\\
21 & 6 497 & 3 &  $\nicefrac{+40}{-20}$ & 14\\
22 & 5 442 & 5 &  $\nicefrac{+22}{-61}$ & 88\\
23 & 4 780 & 7 &  $\nicefrac{+45}{-106}$ & -44\\
24 & 4 229 & 9 &  $\nicefrac{+47}{-281}$ & -142\\
\end{tabular}
\end{ruledtabular}
\caption{Table of measured hyperfine splitting in the range $n=20\ldots24$, as well as fitting error, error derived from piezo nonlinearities and distance to scaling law fit, all given in kHz.}
\label{tab:hf}
\end{table}

The resulting hyperfine splittings are shown in Fig. \ref{fig:scaling}, with our results also listed in table \ref{tab:hf}. The error bars listed in the table are standard errors obtained from the fit. By separately analysing 300 individual traces of the $20s$ measurement we find a mean hyperfine splitting of \unit{7.801}{\mega\hertz} with a standard error of the mean of \unit{7.2}{\kilo\hertz}, i.e. \unit{19}{\kilo\hertz} larger than the results quoted above. As the fitting of the sidebands can be difficult without averaging we consider the values quoted in table \ref{tab:hf} to be more reliable. 

In addition to this it is worth noting here that nonlinearities in the response of the piezo-element used to tune the coupling laser frequency can in principle also skew our results, although this can be minimized by making sure that the observed peaks are not near a turning point of the frequency scan. We estimate the magnitude of this effect by using only either the lower or the upper sidebands for the frequency calibration. We then find a difference of the measured hyperfine splittings for the two cases of between \unit{50}{\kilo\hertz} and \unit{300}{\kilo\hertz}, with the two highest $n$ showing the largest errors, and the three lower $n$ showing errors of less than \unit{80}{\kilo\hertz}. The error bars shown in the figure are based on these results. The nonlinearity in the piezo scan is the biggest uncertainty identified in the frequency calibration.

Our measurements are in excellent agreement with the expected $(n-\delta)^{-3}$ scaling. Here $\delta$ is the quantum defect of the state, depending on both $n$ and $\ell$ and taken from \cite{Li2003}, leading to an effective principal quantum number $n^* = n-\delta$. The inset of Fig. \ref{fig:scaling} shows our data together with earlier results using microwave transitions to other Rydberg states from reference \cite{Li2003} as well as low-n data from reference \cite{Corney1977} after removing the expected $n^*$-scaling.  We see excellent agreement with the low-n data but an offset of approximately $10\%$ compared to the results of \cite{Li2003}. The excellent agreement of our measurements with low-n data might indicate that this offset is due to systematics in the data of \cite{Li2003}. The data of \cite{Li2003} is in agreement with our measurements assuming an error of $1$ in the principal quantum number of their data. An equation for the scaling of the hyperfine splitting based on our data is given in \eqref{eq:scaling}. From this we expect a hyperfine splitting of approximately \unit{80}{\kilo\hertz} at $n=80$. A best fit with variable exponent yields a scaling law with a power of $-2.95(11)$ for our data. 

\beq
\nu_\mathrm{hfs} = \unit{37.1(2)}{\giga\hertz} (n-\delta)^{-3}
\label{eq:scaling}
\eeq

\begin{figure}[htb]
	\centering
	\includegraphics[width=1.00\columnwidth]{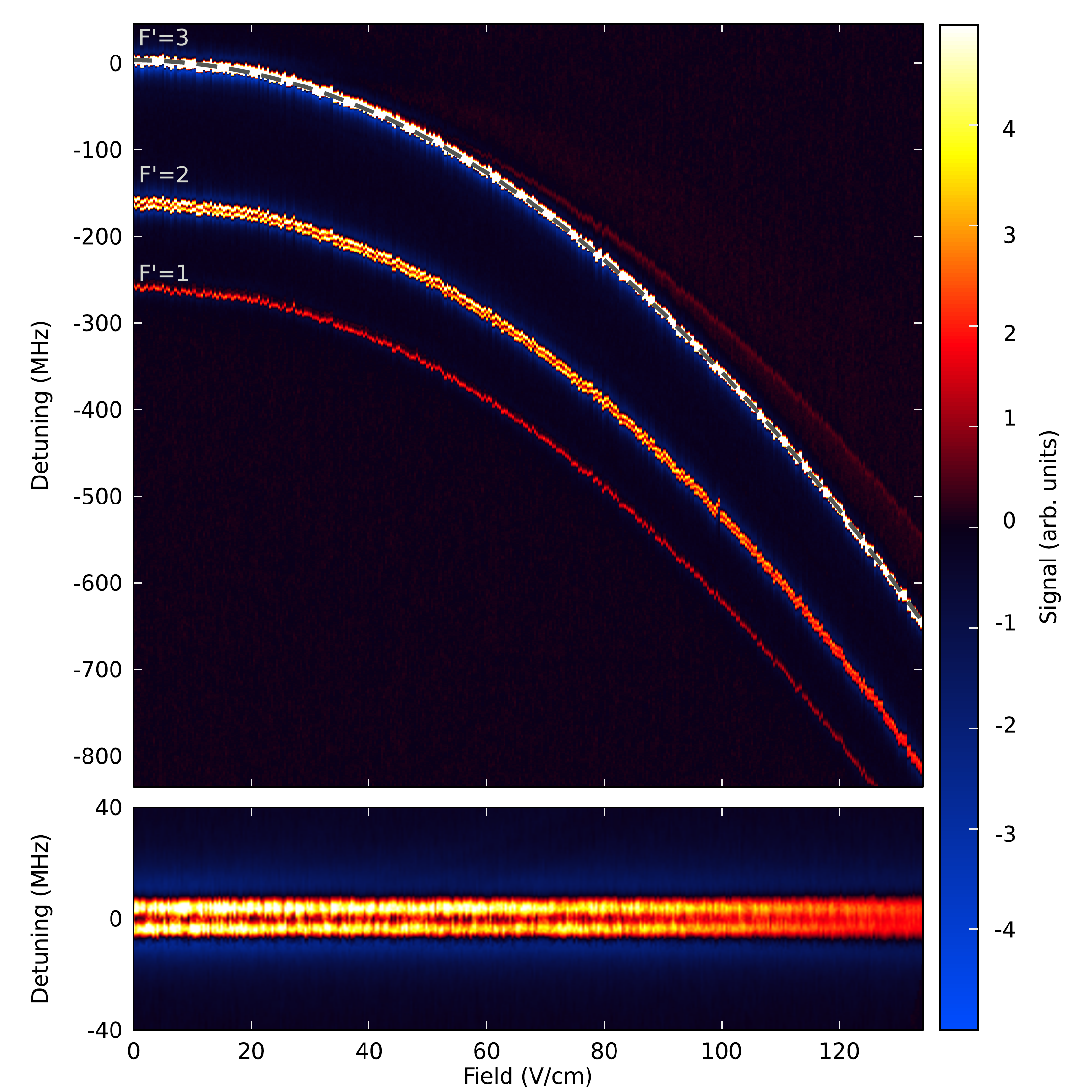}
	\caption{(Color online) Hyperfine-resolved Stark shifts of 20s measured by applying an electric field to the electrodes in the vapour cell. The three lines in the upper part of the figure correspond to three different hyperfine states F$'=1, 2, 3$ in the intermediate $5p_{3/2}$ state. The dashed line overlaid with the F$'=3$ state shows the excellent agreement of the overall Stark shifts with calculations based on the Numerov method \cite{Zimmerman1979}. The bottom part of the figure shows the relative shift of the two hyperfine states after removing the overall Stark-shift, clearly indicating that the hyperfine splitting remains constant in an electric field, and no splitting of $m_F$-components is observed.}
	\label{fig:starkshift}
\end{figure}

%\section{Stark Shift of Hyperfine States}
Finally we present hyperfine-resolved measurements of Stark shifts in fields of up to \unit{130}{\volt\per\centi\meter} for $20s$. The upper part of Fig. \ref{fig:starkshift} shows the overall Stark shift of state $20s$. The three independent lines are due to different $5p_{3/2}$ hyperfine states; while the probe laser is locked to the F$'=2$ transition, other lines can be shifted into resonance by Doppler shift in the vapour cell. Due to the different wavelengths of the probe- and coupling laser these shifts are only partially compensated by the counter-propagating beams; the remaining shifts are expected to be reduced by a factor $\nicefrac{\lambda_c}{\lambda_p} = \nicefrac{480}{780}$ compared to the $5p_{3/2}$ hyperfine splittings, in good agreement with our observations. In the topmost line no hyperfine splitting is visible, as the excitation of the F$''=1$ component of the Rydberg state is dipole forbidden from $5p_{3/2}\ F'=3$. In both the F$'=1$ and F$'=2$ lines the hyperfine splitting of the Rydberg state is in principle visible in the individual traces. However, the signal in F$'=2$ is much stronger than in F$'=1$, making the splitting almost indiscernible for F$'=1$ in this plot.

The overall shift of the Rydberg state is in excellent agreement with calculations based on wavefunctions obtained with the Numerov method \cite{Zimmerman1979}, as can be seen in the dashed line overlaid with the F$'=3$ state which has no free parameters. Fitting a parabola to the Stark shift in Fig. \ref{fig:starkshift} we extract a value of \unit{\alpha = 0.0720 (8)}{\mega\hertz\per(\volt\per\centi\meter)\squared} from this data, in excellent agreement with the theoretically expected value of \unit{\alpha = 0.0722}{\mega\hertz\per(\volt\per\centi\meter)\squared}. The uncertainty in this determination of $\alpha$ is dominated by the accuracy with which the average separation of the electric field plates is known; the uncertainty from the frequency calibration is lower by one order of magnitude. 

We attribute the faint line visible above F$'=3$ to inhomogeneous electric fields at the edges of the cell, in particular in the gap between the electrodes and the cell walls.

The lower part of Fig. \ref{fig:starkshift} shows the hyperfine splitting of the F$'=2$ line after removing the overall quadratic Stark shift of the state. This has been done by fitting the model of eq. \ref{eq:chi} to each individual trace and aligning the center point between the two peaks across all traces. As can clearly be seen no further splitting into $m_F$ sub-levels occurs for these hyperfine states, and the splitting remains constant across the range of fields presented here. This is in agreement with numerical calculations we have performed. At high fields a slight broadening of the peaks can be seen. This is compatible with a misalignment of the field plates by approximately \unit{1}{\milli\rad}, equivalent to \unit{100}{\micro\meter} difference in plate separation at the edges, which we have observed in earlier measurements of higher-lying states in which the hyperfine splitting is entirely negligible.

%\section{Conclusion}
In conclusion we have presented high-precision measurements of the hyperfine splitting of Rydberg states in $^{87}$Rb, achieving \unit{\kilo\hertz} accuracy even in a room-temperature vapour. These measurements obey the expected $(n-\delta)^{-3}$-scaling very well and are in excellent agreement with low-$n$ data such as presented in \cite{Corney1977}. However, our measurements show a small systematic shift compared to measurements at higher $n$ presented in \cite{Li2003}. 

We furthermore present hyperfine-resolved measurements of Rydberg state Stark shifts. To our knowledge no prior measurements of this kind have been demonstrated. These show no change in the  hyperfine splitting as the electric field is increased, and no further splitting of $m_F$ levels, in agreement with our calculations. 

The measurements presented above show how a resolution far below the Doppler limit is possible for Rydberg state spectroscopy in room-temperature vapour cells. Using a vapour cell with internal electrodes as described in this paper this makes high-accuracy Stark spectroscopy extremely simple. This can be of great value for future experiments relying on excellent knowledge of Rydberg-state energies and polarizabilities.

We would like to thank N. J. van Druten for help with the manuscript. This work is part of the research programme of the Foundation for Fundamental Research on Matter (FOM), which is part of the Netherlands Organisation for Scientific Research (NWO). We acknowledge support from the EU Marie Curie ITN COHERENCE Network.

\appendix

\end{document}